\documentclass[prl,aps,10pt,showkeys,nofootinbib,showpacs,twocolumn]{revtex4-1}

\usepackage{amsmath,amssymb,amsfonts,amsthm}

\usepackage{slashed}
\usepackage{mathtools}
\usepackage{amsfonts}
\usepackage{amssymb}
\usepackage{epsfig}
\usepackage{rotating}
\usepackage{url}
\usepackage{times}
\usepackage{color}
\usepackage{bm}
\usepackage{xcolor,colortbl}
\usepackage{hyperref}
\usepackage{enumitem}

\newcommand{\be}{\begin{equation}}
\newcommand{\ee}{\end{equation}}
\newcommand{\bea}{\begin{eqnarray}}
\newcommand{\eea}{\end{eqnarray}}

\newcommand{\plk}{\mathfrak{h}}

\newcommand{\oarX}[1]{\href{http://arxiv.org/abs/#1}{{\ttfamily #1}}}
\newcommand{\arX}[1]{\href{http://arxiv.org/abs/#1}{{\ttfamily arXiv:#1}}}

\newcommand{\dd}{\mathrm{d}}

\def\im{{\rm i}}

\begin{document}

\title{Quantum resolution of the cosmological singularity without new physics}

\author{Steffen Gielen}
\email{s.c.gielen@sheffield.ac.uk}
\affiliation{School of Mathematics and Statistics, University of Sheffield,
Hicks Building, Hounsfield Road, Sheffield S3 7RH, United Kingdom}
\author{Jo\~{a}o Magueijo}
\email{j.magueijo@imperial.ac.uk}
\affiliation{Theoretical Physics Group, The Blackett Laboratory, Imperial College, Prince Consort Rd., London, SW7 2BZ, United Kingdom}

\date{\today}

\begin{abstract}
We study a quantum Hot Big Bang in the connection representation, with a matter constant of motion $m$ whose conjugate defines time. Superpositions in $m$ induce a unitary inner product. The wavefunction reveals a resolution of the singularity problem without new physics or supplementary boundary conditions. Backtracking in time, the probability peak eventually halts at a maximum curvature, its height dropping thereafter while a symmetric contracting peak rises. The Big Bang is replaced by a superposition 
of contracting and expanding regular Universes. We contrast these findings with the situation in the metric representation, where boundary conditions at the singularity are needed for unitary evolution.
\end{abstract}

\maketitle

\section{Introduction}

Spacetime singularities are among the most troubling features of classical General Relativity (GR). In particular, the standard Big Bang model of cosmology implies that our Universe began at a singularity, a rather unsatisfactory situation. In this Letter, we consider the simplest approximation to the early Universe: 
dominated by radiation and homogeneous, isotropic and spatially flat. In such a model any observer would have encountered a singularity at a finite time in the past \cite{HawkingEllis}.

A possible alternative to the singularity is a quantum ``bounce'' from a contracting into our current expanding phase. Most bounce models rely on physics beyond GR, such as string theory or loop quantum gravity \cite{BounceReview} (see also \cite{MatrixCosmo} for a recent proposal). Here we will instead revisit an older question: are the principles of quantum mechanics and GR sufficient to resolve the Big Bang singularity? This is a long-standing question in quantum cosmology going back over 50 years \cite{misner}. There is no consensus on its answer, with ambiguities both in the definition of quantum cosmology models and in the criteria for singularity resolution \cite{QSingRes}.

The classical history of the Universe has only had a finite time in the past, but standard unitarity demands that a quantum state can be translated arbitrarily far into the past by a time-evolution operator. This already seems to imply that the quantum evolution must deviate from the classical Big Bang, making it non-singular. There are, however, many subtleties. Foremost, 
the ``timelessness'' of the Wheeler--DeWitt equation forces us to define ``quantum time'' in a relational sense from the dynamical degrees of freedom \cite{ProblemofTime}. The relational time distance to the Big Bang may or may not be finite.  If it is, one would expect singularity resolution, whereas if it is not, the quantum theory could still be singular \cite{GotayDemaret}. But unitarity itself may require 
postulating boundary conditions (e.g., \cite{GielenMenendez}): one could argue that this simply reverse-engineers a solution.

In this Letter we add a significant twist: we examine the problem in the connection representation. 
The Big Bang singularity occurs at zero scale factor $a$  (metric), and at infinite extrinsic curvature (connection). 
The difference might seem innocuous, but the quantum theories and their solutions are radically different. In the metric representation
we must add reflecting boundary conditions at $a=0$ to enforce unitarity for $a\ge 0$, and create a bounce. 
In contrast, the simplest unitary theory in the connection representation displays a probability peak that, backtracking in time, eventually halts at a maximum curvature, 
its height dropping thereafter while a symmetric contracting peak rises, and eventually connects to a semiclassical contracting Universe.  
The Big Bang itself is replaced by a superposition 
of contracting and expanding regular Universes, without the need for any boundary condition.

\section{Theoretical tools}

Cosmological models contain a number of dynamically conserved quantities $\alpha_i$: for instance, for a perfect fluid such a quantity arises from the continuity equation. Such quantities also appear in approaches that elevate fundamental parameters of Nature to dynamical quantities subject to a conservation law. This is how the cosmological constant $\Lambda$ becomes an integration constant in unimodular gravity~\cite{unimod1,unimod}, but one can apply similar ideas to Newton's constant, spatial curvature, 
or the Planck mass~\cite{pad1,vikman,TimeConstants,BrunoJoao,JoaoPaper}. Importantly, each conserved quantity can play the role of ``energy'', and its conjugate variable that of time.

Concretely, if $q_A$ are other degrees of freedom of geometry and matter, the Hamiltonian constraint can either be written in terms of $\alpha_i$ (resulting in the 
standard Wheeler--DeWitt equation for timeless $\psi_s(q_A,\alpha_i)$) or in terms of their conjugate ``times'' $T_i$ 
(leading to a Schr\"odinger-like equation for $\psi(q_A,T_i)$)~\cite{GielenMenendez,TimeConstants,JoaoPaper}. 
The simplest case is that of a single $\alpha$ and single $q$, such that in a preferred operator ordering the general solution is
\be\label{gensol}
\psi(q,T)  =\int \frac{\dd\alpha}{\sqrt {2\pi\plk }} {\cal A}(\alpha) \exp{\left[\frac{\im }{\plk}\alpha (X(q)  - T)  \right]}
\ee
with suitably chosen function $X(q)$ and ``effective Planck constant'' $\plk$, to be defined shortly. In these cases minisuperspace behaves like a dispersive medium~\cite{JoaoPaper}, with packets changing their shape (in $q$) as they propagate. $\alpha$, $X$ and $T$ are the medium's linearizing variables: they remove
 dispersion when waves are written in terms of them~\cite{DSR,JoaoPaper}.

A key strength of this approach which we will exploit here is that the unitarity of the quantum theory is guaranteed: one can define an inner product in terms of the amplitudes~\cite{JoaoPaper,bbounce},
\bea\label{innalpha}
\langle\psi_1|\psi_2  \rangle  &=&\int \dd\alpha \; {\cal A}_1^\star (\alpha) {\cal A}_2(\alpha)\,,
\eea
which is automatically conserved. For (\ref{gensol}), after changing variables from $q$ to $X$, (\ref{innalpha}) can be written as
\be\label{innX}
\langle\psi_1|\psi_2  \rangle=\int {\rm d}X \psi_1^\star(X,T)\psi_2(X,T)
\ee
if $\alpha,X$ vary over all of $\mathbb{R}$, which is a condition for the unitarity of (\ref{innX}). 
Note that whereas replacing the conserved $\alpha$ by a nontrivial function $\beta=\beta(\alpha)$ leads to  {\em classically}
equivalent theories (with new conjugate $T_\beta=T_\alpha/\beta'(\alpha)$), 
the corresponding quantum theories are not equivalent. Their inner products  are different, since
${\cal A}(\alpha){\rm d}\alpha={\cal A}(\beta)\dd\beta$ implies
\be
\int \dd\alpha \; {\cal A}_1^\star (\alpha) {\cal A}_2(\alpha)=\int \dd\beta \; {\cal A}_1^\star (\beta) {\cal A}_2(\beta)\beta'(\alpha)\,.
\label{innalphabeta}
\ee
Their time evolution is also different, since generally $\alpha \cdot T_\alpha\neq \beta \cdot T_\beta$; cf.~Eq.~(\ref{gensol}).
Finally, states coherent in $\alpha$ (with Gaussian ${\cal A}(\alpha)$) are generally not coherent in $\beta$. 

We now specify a particular cosmological model.
At high energies, matter has equation of state $w=1/3$.  Our action for GR with matter is 
\be
S = \frac{3V_c}{8\pi G}\int {\rm d}t \left(\dot{b}a^2+\dot{m}T- N a\left(-(b^2+k) +\frac{m}{a^2}\right)\right)
\label{GRmatter}
\ee
where 
$b$ represents the connection (on-shell the inverse Hubble length, $b=\dot{a}/N$),
$k$ is spatial curvature, and $m$ is a quantity associated to matter whose conservation is enforced by the second term. $V_c$ is the coordinate volume of space. Variation with respect to the lapse $N$ enforces the standard Hamiltonian constraint
\be\label{ham0}
-(b^2+k) a^2+m= 0\,.
\ee
Below we comment on how the $m$ terms in Eq.~(\ref{GRmatter}) can be derived from different starting points, but what follows does not depend on these discussions. 

\section{Singularity resolution in the connection representation} 

Action (\ref{GRmatter}) suggests canonical pairs
\be
\{b,a^2\} = \{m,T\}=\frac{8\pi G}{3V_c}\, ,
\ee
so that upon quantization 
$\left[b,a^2\right]=\left[m,T\right]=\im \plk:=\frac{8\pi\im G \hbar}{3 V_c}$.
In the ordering required in the general formalism introduced above, Eq.~(\ref{ham0}) becomes
\be
\im\plk\frac{\partial}{\partial T}\,\psi(b,T)=-\im \plk (b^2+k)  \frac{\partial}{\partial b}\psi(b,T)\,,
\label{wdw}
\ee
i.e., the promised Schr\"odinger equation. As an aside, we stress that while $b$ and $a^2$ are analogous to the connection variable $c$ and densitised triad $p$ appearing in loop quantum cosmology \cite{LQCreview}, we apply a standard (Wheeler--DeWitt) canonical quantization without including any putative quantum geometry effects. Our notion of unitarity based on a matter clock also differs from what is done in loop quantum cosmology.

This ordering implies that the solutions are indeed of the form (\ref{gensol}), with 
\be
\alpha_b=m\label{alpham0}\,,\quad X_b=\int^b \frac{{\rm d}\tilde{b}}{(\tilde{b}^2+k)}
\ee
or $X_b=-1/b$ for $k=0$, which we now assume. 
These solutions are an adaptation for radiation of a generalized Chern--Simons state for $\Lambda$ \cite{JoaoPaper}. The inner product (\ref{innX}), which takes the explicit form
\be
\langle\psi_1|\psi_2  \rangle=\int \frac{{\rm d}b}{b^2} \psi_1^\star(b,T)\psi_2(b,T)\,,
\label{innerb}
\ee
is conserved in $T$, and the gravitational part $-\im \plk\, b^2\frac{\partial}{\partial b}$ of the Hamiltonian constraint is self-adjoint. Notice that this quantization leads to operators representing the classical quantities $m$ and $a^2$ with positive and negative eigenvalues. This is as it must be, given that the quantity $\alpha$ always has to vary over the real line for the theory to be unitary. This feature implies an extension of the phase space of the classical theory (where one assumes $a^2\ge 0$) in the corresponding quantum theory, which is crucial for preserving unitarity: states of negative $a^2$ must in general be allowed to contribute. Indeed, both here and in the different theory studied below, restricting the dynamical variables to enforce a closer correspondence to the classical starting point makes unitarity more difficult to achieve.

One can interpret states with negative $a^2$ in terms of Euclidean geometries, given the signature change in the corresponding spacetime metric. This interpretation is consistent with the fact that in Euclidean signature the constraint (\ref{ham0}) (with $k=0$) becomes
\be
b_E^2 a_E^2+m= 0
\ee
for a Euclidean connection $b_E$ and scale factor $a_E$ such that $b_E$ is canonically conjugate to $a_E^2$. We will only be interested in states sharply peaked around some $m_0>0$, so that these negative eigenvalues do not contribute and their physical interpretation is not important for what follows.

More concretely, different choices of states are possible, with some disabusing the constant $m$ of its 
name, but our Universe happens to have reasonably sharp ``constants''; 
this is related to the existence of a semiclassical limit at late times.
We choose ${\cal A}(m)=\sqrt{{\bf N}(m_0,\sigma_m)}$ where ${\bf N}(m_0,\sigma_m)$ is a (normalized) normal distribution with mean $m_0$ and standard deviation $\sigma_m$. Integrating (\ref{gensol})
we obtain a squeezed-coherent state in $X_b$ (not in $b$, we stress),
\bea\label{coherent}
\psi(b,T)&=&
e^{\frac{\im}{\plk}m_0(X_b-T)}
\frac{\exp\left[-\frac{(X_b -T )^2}{4\sigma_T^2}\right]}{(2\pi \sigma_T ^2)^{1/4}}
\eea
with $\sigma_T = \plk/2\sigma_m$ saturating the Heisenberg relation. Using Eq.~(\ref{innerb})  we find
the probability for a given $b$ at time $T$:
\be\label{ProbbT}
{\cal P}(b,T)=\frac{1}{b^2}\frac{\exp\left[-\frac{\left(\frac{1}{b}+T\right)^2}{2\sigma^2_T}\right]}{\sqrt{2\pi \sigma^2_T}}\,.
\ee
Unitarity is manifest as the statement $\int {\rm d}b\;{\cal P}(b,T)=1$ for all $T$.
While $\sigma_T$ is a free parameter (which one may think of as the analogue of the Planck time, giving a fundamental uncertainty to the concept of time), most importantly it is constant in $T$. Classically
$\dot T=-\frac{N}{a}$
so, within the conditions of Ehrenfest's theorem, $\langle T\rangle $ is minus conformal time $\eta$.

\begin{figure}
\includegraphics[scale=.9]{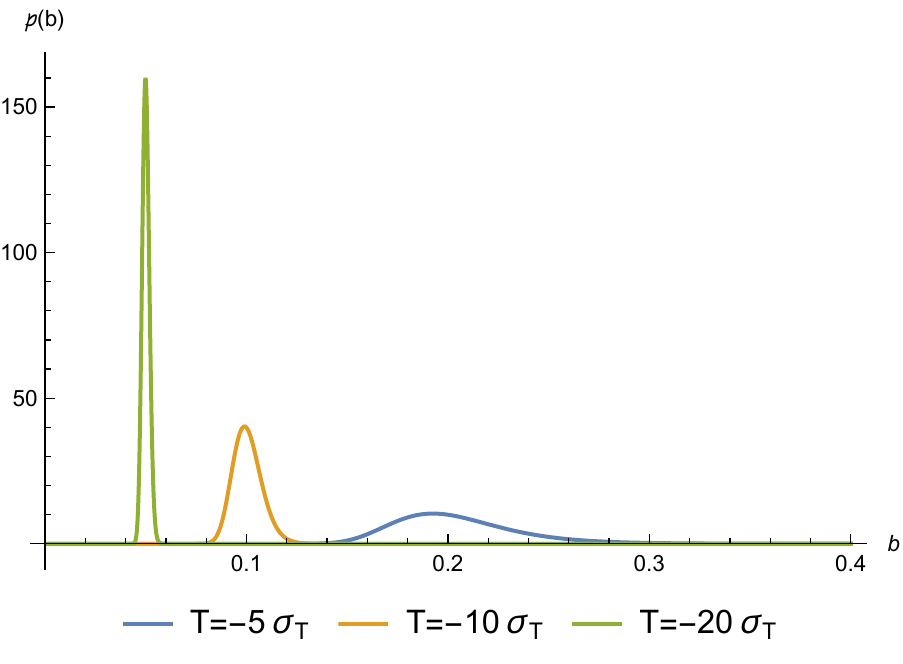}
\caption{As $|T|\gg \sigma_T$ the distribution ${\cal P}(b)$ quickly becomes near-Gaussian in $b$, with $\sigma(b)/b\ll 1$. We can identify $T=-\eta$ since $\sigma(T)/|T|\ll 1$, so that in the expanding branch ($T<0$, $\eta>0$) the ever-sharper peak follows the classical trajectory $b=1/\eta$.}
\label{SuperPlanck-prob}
\end{figure}

Eq.~(\ref{ProbbT}) illustrates the dispersive nature of the medium, essential for our solution of the singularity 
problem. Wave packets sharpen up for increasing $|T|\gg \sigma_T$, but lose their WKB shape and 
become fully quantum as $|T|\lesssim \sigma_T$. We display the first (semiclassical) behavior in Fig.~\ref{SuperPlanck-prob} for an 
expanding Universe ($T<0$, $\eta>0$). We can use the relation $T= -\eta$ because
$\sigma(T)/|T|=\sigma_T/|T|\ll 1$. The peak moves along the classical trajectory $b=1/\eta$
with ever tinier fractional standard deviation $\sigma(b)/|b|$
(since $\sigma_T=\sigma(X_b)$ implies $\sigma_b\approx \sigma(X_b)/|X'_b|=b^2\sigma_T$).
The state therefore is near-coherent (indeed delta-like) in $b$:
\be
{\cal P}(b,\eta)\approx \frac{\exp\left[-\frac{\left(b-\frac{1}{\eta}\right)^2}{2\sigma^2_b }\right]}{\sqrt{2\pi \sigma^2_b}}\,,\quad \sigma_b\approx
\sigma_T/\eta^2\,.
\ee

\begin{figure}
\includegraphics[scale=.9]{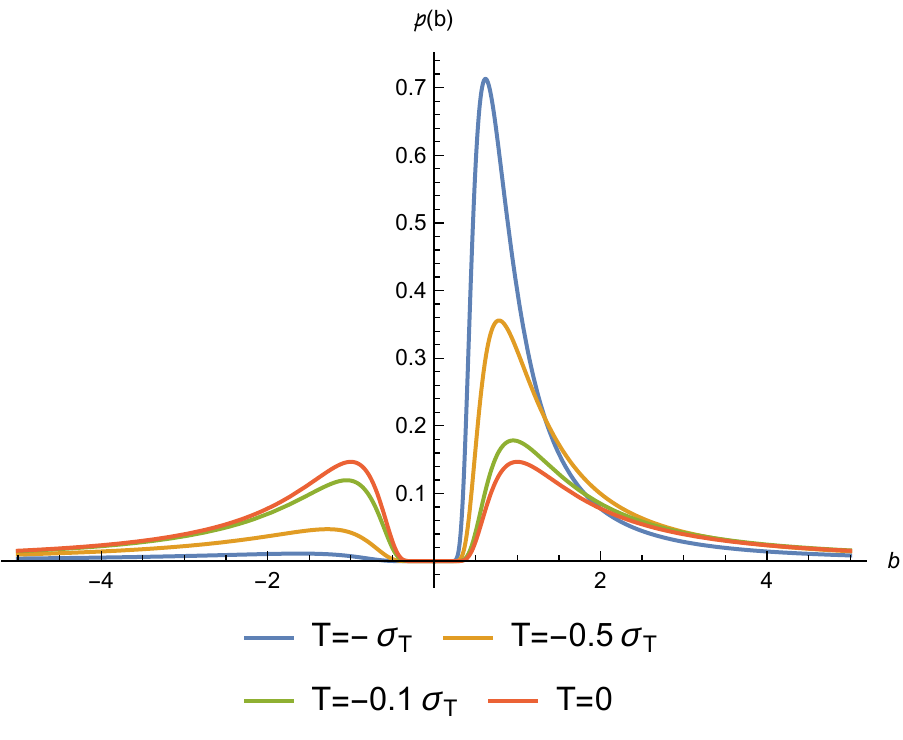}
\caption{For $|T|\lesssim \sigma_T$ the distribution ${\cal P}(b)$ is very distorted, and its peak does not go to infinity but
saturates at $b=b_P$. As $T\rightarrow 0^-$ this peak lowers, and a secondary peak in the contracting zone becomes more prominent. At $T=0$ the two peaks have the same height, but nothing is singular. For $T>0$
a symmetric film is played, eventually linking up to a semiclassical contracting phase [see animation in Supplementary Material]. We thus have a quantum bounce.}
\label{Planck-prob}
\end{figure}

In contrast, for $|T|\lesssim  \sigma_T$ the dispersive nature of the medium is all-important, as shown in 
Fig.~\ref{Planck-prob}. The packet widens and becomes grossly distorted, disallowing a WKB approximation. 
In addition the probability peak gets stuck at $b\approx  b_P=1/(\sqrt{2}\sigma_T)
=\sqrt 2\sigma_m/\plk$,
instead of going to infinity, as expected from the singular classical trajectory.\footnote{Notice that this effective curvature limit $b_P$ depends on the chosen state, rather than being a fundamental parameter in the theory.}  The distribution is very skewed, 
with a tail that is  more prominent for smaller $|T|$, whilst the height of the peak decreases. Meanwhile,
a contracting peak emerges at $b\approx - b_P<0$, its height rising in tandem with the first peak's dropping. 
The wavefunction is always regular, and indeed at  $T=0$ it is symmetric with
\be
{\cal P}(b,T=0) =  \frac{1}{b^2}\frac{\exp\left[-\frac{\left(\frac{1}{b}\right)^2}{2\sigma^2_T}\right]}{\sqrt{2\pi \sigma^2_T}}\,.
\ee
The ``Big Bang'' is therefore replaced by
a perfectly balanced quantum superposition of a contracting and expanding Universe.  
For $T>0$ the peak at $b=-b_P$ supersedes that at $b=b_P$ and continues growing until 
for $T\sim \sigma_T$  it starts to move towards smaller $|b|$, linking up 
with a semiclassical contracting  Universe when $T\gg \sigma_T$.
These results can be derived directly from Eq.~(\ref{ProbbT}) and they 
are due to the form of $\psi$ (and $X(b)$) and the measure ${\rm d}\mu(b)$.

We therefore have a quantum model for a nonsingular bouncing Universe. It avoids the singularity because time around the classical singularity is smeared by quantum uncertainty. It results directly from the fact that unitarity requires $X(b)$ {\it not} to be constrained (it must cover the whole real line, so that (\ref{innX}) is conserved): $b$ should not be constrained to an expanding Universe. Unitarity {\it forces} a bounce directly, precisely by ruling out any constraints on $b$ and the need for any associated boundary conditions. 
This is the radical implication of our Letter, to be contrasted with the picture to emerge from the more familiar metric formulation, as we now show.

\section{Singularity resolution in metric variables} 

The metric representation is based on the canonical pair $\{a,p_a\} = \frac{8\pi G}{3V_c}$ with $p_a:=-2ba$.
The constraint (\ref{ham0}) becomes $m=\frac{1}{4}p_a^2$ leading to the Schr\"odinger equation
\be
\im\plk\frac{\partial}{\partial T}\,\psi(a,T)=-\frac{1}{4}\plk^2\frac{\partial^2}{\partial a^2}\psi(a,T)\,.
\label{wdw2}
\ee
This fits into the formalism based on Eq.~(\ref{gensol}), with linearizing variables $X_a=a$, $\alpha_a$ satisfying $\alpha_a^2 = 4m$  (where $m\ge 0$ but $\alpha_a$ is unrestricted) and $T_a=T/\alpha_a'(m)$. Eq.~(\ref{innX}) is then simply
\be
\langle\psi_1|\psi_2\rangle = \int {\rm d}a\;\psi_1^\star(a,T)\psi_2(a,T)\,.
\label{L2innerprod}
\ee
This inner product differs from the one in the connection representation since $\alpha_a=\pm 2 \sqrt{m}$ and $\alpha_b=m$ are different: these are two different quantum theories. Another difference between the theories is that $m$ is now restricted to be positive, even though the new $\alpha$ can still take any real value (again, as it must).

A priori the scale factor $a$ could take any real value, but if we restrict $a\ge 0$, unitarity is no longer guaranteed; the operator on the right-hand side of (\ref{wdw2}) is no longer self-adjoint~\cite{selfadj}. Instead, demanding $\langle\psi_1|\frac{\partial^2}{\partial a^2}\psi_2\rangle=\langle\frac{\partial^2}{\partial a^2}\psi_1|\psi_2\rangle$ on $[0,\infty)$ requires a Robin boundary condition
\be
\frac{\partial\psi}{\partial a}(0,T)=\gamma\psi(0,T)
\label{boundcond}
\ee
where $\gamma$ is a free parameter (which can be $\infty$). The operator $\partial^2/\partial a^2$ has a one-parameter family of self-adjoint extensions.

Unitarity then leads to a reflecting boundary condition: rather than ``disappearing'' through $a=0$ the quantum state is reflected back to positive $a$. Inserting (\ref{gensol}) into (\ref{boundcond}) leads to
\be\label{AMconst}
{\cal A}(\pm |\alpha_a|)={\cal B}(|\alpha_a|)\left(1\mp \im\frac{\gamma\plk}{|\alpha_a|}\right)
\ee
for some function ${\cal B}$ on the positive half-line. Eq.~(\ref{gensol}) becomes
\bea
\psi(a,T)&=&\int_0^{\infty} \frac{{\rm d}m}{\sqrt{2\pi\plk}}\,\mathcal{C}(m) \,\exp\left(-\frac{\im}{\plk}m T\right)\label{wf1}
\\&& \times\left(\cos\left(\frac{2 \sqrt{m} a}{\plk}\right)+\frac{\gamma\plk}{2 \sqrt{m}}\sin\left(\frac{2 \sqrt{m} a}{\plk}\right)\right)
\nonumber
\eea
where $\mathcal{C}(m)=2\mathcal{B}(m)/\sqrt{m}$, and Eqs.~(\ref{innalpha})--(\ref{innX}) now
become the time-independent
\be\label{metricnorm}
\int_0^\infty {\rm d}a\;|\psi|^2 = \int_0^\infty {\rm d}m\,|\mathcal{C}(m)|^2\left(\frac{\sqrt{m}}{4}+\frac{\gamma^2\plk^2}{16\sqrt{m}}\right).
\ee
The theory is now unitary at the cost of introducing boundary conditions dependent on the parameter $\gamma$.

\begin{figure}
\includegraphics[scale=.85]{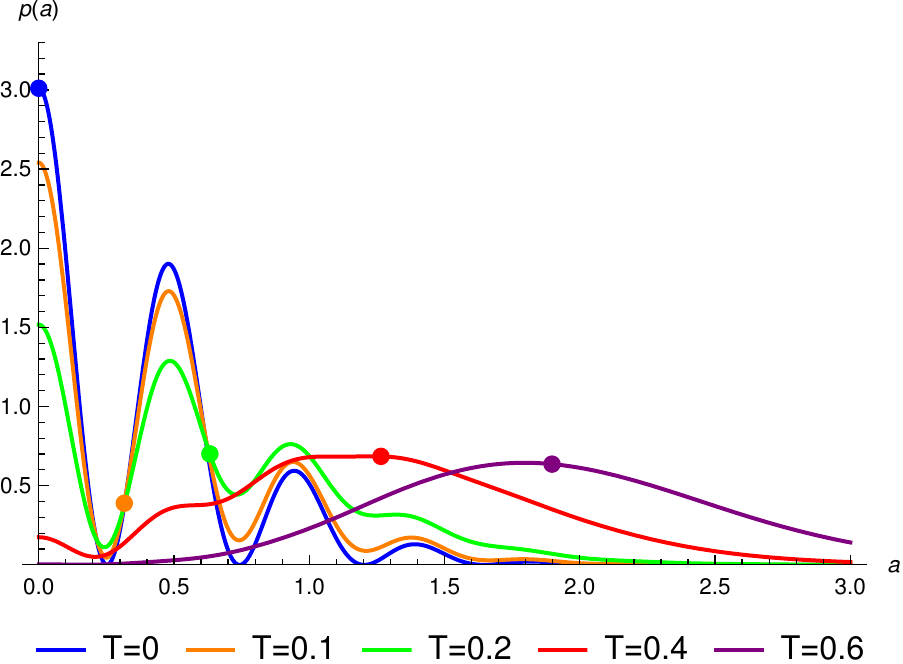}
\caption{Probability distribution $p(a,T)=|\psi(a,T)|^2$ for different values of $T$, for $m_0=10$ and $\sigma_m=3$. Dots show the classical solution  $a(T)=\sqrt{m_0} T$ corresponding to these values of $T$. Again, everything is symmetric under $T\rightarrow -T$. At small $|T|$ expectation values depart from classical values due to the non-classical peaks. As $T$ moves away from the classical singularity we converge to the classical solution.}
\label{fig2}
\end{figure}

Choosing again ${\cal C}(m)=\sqrt{{\bf N}(m_0,\sigma_m)}$, in Fig.~\ref{fig2}
we show how the boundary condition leads to quantum departures from the classical solution. At small $|T|$ there is interference between a classical contracting and a classical expanding solution, which leads to the presence of multiple peaks at different finite values of $a$. The expectation value $\langle a(T)\rangle$ deviates from the classical solution in this region, and is bounded away from zero indicating singularity resolution. Fluctuations over the expectation value, however, are large.
Again this behavior follows from the quantum nature of time for small $|T|$, together with unitary dynamics. But by enforcing unitarity through boundary conditions (explicitly eliminating $a=0$) one has effectively reverse-engineered this solution, in contrast with the more immediate results in the connection representation.

With some analogy to what was done in the connection representation, we might propose an extension of the classical phase space in the quantum theory, in which we let $a$ take either positive or negative values. Since only $a^2$ enters the metric, there is no physical observable that could detect the sign of $a$, so this would amount to gluing the original phase space to its mirror image. In this case, no boundary conditions would be needed as the operator $\frac{\partial^2}{\partial a^2}$ is already self-adjoint. Our cosmological model would now correspond to a quantum particle on the real line, which happily passes through $a=0$; there would be no notion of singularity resolution in this theory, in contrast to what we have observed in the connection representation.

\section{Discussion} 

Our results show that unitary time evolution in quantum cosmology leads to a resolution of the singularity, without the 
need for boundary conditions in the connection representation. This happens in part 
 because  the time variable $T$ is conjugate to a classical constant of motion that is not infinitely sharp, but also not entirely undefined. Hence, $T$ has a fixed uncertainty $\sigma_T$.  As we plunge into the classical singularity at $T=0$, eventually $|T|\sim\sigma_T$ so that quantum fluctuations become significant, leading to deviations from the classical trajectory. In the connection representation the spread in $T$ is translated into a smearing of $b$ from $b_P$ (where the probability peak gets stuck) to infinity. The probability of infinite $b$ is always zero, even as $T\rightarrow 0$. The wavefunction then develops non-negligible support at $b<0$, so that $|b|$ and its sign are undefined: at the ``Big Bang'' the Universe is in a superposition of contracting and expanding phases.  
Unlike in the metric representation, there is no reflection or interference (``ringing''), simply a regular quantum transition through the classical Big Bang, transferring the probability peak from contraction to expansion at finite curvature [see Supplementary Material for animation]. There is no need to excise a region and impose boundary  conditions. 

An important point in our analysis concerns the extension of the classical gravitational phase space (described by a real $b$ and positive definite $a^2$) in the quantum theory, implying that in the connection representation the variable $a^2$ can take negative values. We interpreted these new configurations as Euclidean. This extension is essential for obtaining a theory that is both unitary and resolves the singularity; understanding its interpretation beyond minisuperspace could give important insights into the nature of quantum gravity. In metric variables, where we have a real-valued scale factor $a$ (perhaps extended to negative values), we do not have access to this extended phase space. Hence there can be no unitary mapping between the two theories we discussed, or from the connection representation to another theory written in terms of $a$.  At best, the theory obtained starting from the $b$
representation can be transposed to a representation diagonalizing an unrestricted $a^2$, using the Fourier transform which for a pure Lambda relates the Chern--Simons--Kodama state and the Hartle--Hawking wave function \cite{BrunoHH}. In that case, the transformation is unitary if we transform the inner product in the $b$ representation appropriately to the $a^2$ representation. The same transformation is more intricate for pure radiation, where it is likely to lead to a nonlocal inner product.

Given that we are dealing with the Planck epoch, one may wonder how fundamental our theory is. 
Several theories lead to Eq.~(\ref{GRmatter}), some blatantly ``effective'', others with ``fundamental'' pretensions, but all
well-defined beyond minisuperspace.
We may frame our model as a perfect fluid, with action
\be
S_{{\rm fl}} = \int {\rm d}^4 x \left[-\sqrt{-g}\,\rho\left(\frac{|J|}{\sqrt{-g}}\right)+J^\mu\left(\partial_\mu\varphi+\beta_A\partial_\mu\alpha^A\right)\right]
\nonumber
\ee
where $J^\mu$ is a vector density representing the densitized particle number flux, $|J|=\sqrt{-g_{\mu\nu}J^\mu J^\nu}$ and $\varphi$, $\beta_A$ and $\alpha^A$ are suitable Lagrange multipliers~\cite{Brown}.
Choosing the appropriate function $\rho$ for radiation, $\rho(n)\propto n^{4/3}$, and reducing the action to minisuperspace would lead to Eq.~(\ref{GRmatter}), as in~\cite{Brown2}.
This is conservative, but conversely one may question the validity of using a perfect fluid description
in the Planckian regime. 

Alternatively, we may derive (\ref{GRmatter})  from a theory of constants of Nature~\cite{pad1,vikman,TimeConstants,JoaoPaper}
carbon-copied from the covariant formulation of unimodular gravity~\cite{unimod}.
In such theories, after choosing a  constant $\alpha$, one replaces the standard action
$S_0$ by
\be\label{Utrick}
S=
S_0+ \int {\rm d}^4 x\, (\partial_\mu\alpha)  T^\mu_\alpha\,,
\ee
where $\alpha$ is a scalar and $T_\alpha^\mu$ again a vector density, so the 
added term is diffeomorphism invariant. Then, $\alpha$ becomes a constant-on-shell-only  ($\partial_\mu \alpha=0$
is an equation of motion),  with conjugate ``time'' $T_\alpha^0$ (for $\Lambda$, this time is proportional to the spacetime volume to the observer's past). 
Applications of this approach may be found in~\cite{pad1} for the Planck mass, and~\cite{vikman,BrunoJoao} for the gravitational coupling. The latter would lead to Eq.~(\ref{GRmatter}).  All these approaches would lead to the results we have presented, which are blind to their roots and so model-independent.

Our model is simple, but captures the most relevant degrees of freedom of the radiation era in the very early Universe. One may ask what happens when we include additional degrees of freedom, in particular anisotropies that would be expected to {\it classically} dominate at the very earliest times. This is a worry in all bounce scenarios \cite{BounceReview} subject to this classical instability, but we stress that this 
is avoided in any quantum approach where the curvature remains ``stuck'' at a finite maximum, as in our connection quantization. In this case, all one would need to ensure is that anisotropies do not yet dominate before the deep quantum regime is reached, which is the case for a wide range of initial conditions. In the metric representation, Ref.~\cite{GielenMenendez} included a massless scalar field and found singularity resolution similar to our results. Since a massless scalar field has energy density $\sim a^{-6}$, this result would carry over to anisotropies. Our results indicate that it is really unitarity with respect to a suitable matter clock, and the uncertainty in the clock, that are responsible for quantum singularity resolution, rather than any model-specific assumptions. 
Studying an anisotropic model in more detail could be useful to understand whether other quantities, such as directional Hubble rates, also necessarily remain bounded if the overall (mean) Hubble rate does.

\section*{Acknowledgments}

We thank Bruno Alexandre, Robert Brandenberger, Gabriel Herczeg and Lee Smolin for discussions. This work was supported by a Royal Society University Research Fellowship (UF160622) and Research Grant RGF$\backslash$R1$\backslash$180030 (both SG) and by the STFC Consolidated Grant ST/T000791/1 (JM).

\end{document}